\documentclass[11pt]{article}
\pdfoutput=1
\usepackage{fullpage}
\usepackage{cite}
\usepackage{graphicx}
\usepackage{amsmath}
\usepackage{amssymb}
\usepackage{subcaption}
\title{}
\date{}
\renewcommand{\vec}[1]{\mbox{\boldmath$ #1 $}}

\def\beq{\begin{equation}}
\def\eeq{\end{equation}}
\allowdisplaybreaks
\begin{document}
\bibliographystyle{utphys}
\newcommand{\msbar}{\ensuremath{\overline{\text{MS}}}}
\newcommand{\DIS}{\ensuremath{\text{DIS}}}
\newcommand{\abar}{\ensuremath{\bar{\alpha}_S}}
\newcommand{\bb}{\ensuremath{\bar{\beta}_0}}
\newcommand{\rc}{\ensuremath{r_{\text{cut}}}}
\newcommand{\Nd}{\ensuremath{N_{\text{d.o.f.}}}}
\setlength{\parindent}{0pt}

\titlepage
\begin{flushright}
QMUL-PH-20-01\\
\end{flushright}

\vspace*{0.5cm}

\begin{center}
{\bf \Large Monopoles, shockwaves and the classical double copy}

\vspace*{1cm} 
\textsc{Nadia Bahjat-Abbas\footnote{n.bahjat-abbas@qmul.ac.uk},
  Ricardo Stark-Much\~{a}o\footnote{r.j.stark-muchao@qmul.ac.uk}, 
 and Chris D. White\footnote{christopher.white@qmul.ac.uk}} \\

\vspace*{0.5cm} Centre for Research in String Theory, School of
Physics and Astronomy, \\
Queen Mary University of London, 327 Mile End
Road, London E1 4NS, UK\\

\end{center}

\vspace*{0.5cm}

\begin{abstract}
The classical double copy relates exact solutions in biadjoint scalar,
gauge and gravity theories. Recently, nonperturbative solutions have
been found in biadjoint theory, which have been speculated to be
related to the Wu-Yang monopole in gauge theory. We show that this
seems not to be the case, by considering monopole solutions in the
infinitely boosted (shockwave) limit. Furthermore, we show that the
Wu-Yang monopole is instead related to the Taub-NUT solution, whose
previously noted single copy is that of an abelian-like (Dirac)
monopole. Our results demonstrate how abelian and non-abelian gauge
theory objects can be associated with the same gravity object, and
clarify a number of open questions concerning the scope of the
classical double copy.
\end{abstract}

\vspace*{0.5cm}

\section{Introduction}
\label{sec:intro}

Field theories continue to be relevant in many different areas of
physics. Of particular interest are relativistic quantum theories,
needed for particle physics and gravity. Studying the relationships
between different theories can be just as important as examining
individual theories themselves, given that this may reveal new
conceptual insights, or computational methods. One such relationship
is the {\it double copy}~\cite{Bern:2008qj,Bern:2010ue,Bern:2010yg},
whose original incarnation related scattering amplitudes in
non-abelian gauge, and gravity, theories - including their
supersymmetric generalisations (see
refs.~\cite{Bern:2010ue,Bern:1998ug,Green:1982sw,Bern:1997nh,Carrasco:2011mn,Carrasco:2012ca,Mafra:2012kh,Boels:2013bi,Bjerrum-Bohr:2013iza,Bern:2013yya,Bern:2013qca,Nohle:2013bfa,
  Bern:2013uka,Naculich:2013xa,Du:2014uua,Mafra:2014gja,Bern:2014sna,
  Mafra:2015mja,He:2015wgf,Bern:2015ooa,
  Mogull:2015adi,Chiodaroli:2015rdg,Bern:2017ucb,Johansson:2015oia,Oxburgh:2012zr,Geyer:2019hnn,White:2011yy,Melville:2013qca,Luna:2016idw,Saotome:2012vy,Vera:2012ds,Johansson:2013nsa,Johansson:2013aca,Bargheer:2012gv,Huang:2012wr,Chen:2013fya,Chiodaroli:2013upa,Johansson:2014zca,Johansson:2017srf,Chiodaroli:2017ehv,Chen:2019ywi,Plefka:2019wyg,Aoude:2019xuz,Lipstein:2019mpu},
and ref.~\cite{Bern:2019prr} for a comprehensive review). This was
subsequently extended to classical
solutions~\cite{Monteiro:2014cda,Luna:2015paa,Luna:2016due,Goldberger:2016iau,Anastasiou:2014qba,Borsten:2015pla,Anastasiou:2016csv,Anastasiou:2017nsz,Cardoso:2016ngt,Borsten:2017jpt,Anastasiou:2017taf,Anastasiou:2018rdx,LopesCardoso:2018xes,Goldberger:2017frp,Goldberger:2017vcg,Goldberger:2017ogt,Luna:2016hge,Luna:2017dtq,Shen:2018ebu,Levi:2018nxp,Plefka:2018dpa,Cheung:2018wkq,Carrillo-Gonzalez:2018pjk,Monteiro:2018xev,Plefka:2019hmz,Maybee:2019jus,Johansson:2019dnu,PV:2019uuv,Carrillo-Gonzalez:2019aao,Bautista:2019evw,Moynihan:2019bor,Bah:2019sda,CarrilloGonzalez:2019gof,Goldberger:2019xef,Kim:2019jwm,Banerjee:2019saj},
which has a number of applications. Firstly, there is the possibility
that the double copy could greatly streamline calculations in
classical General Relativity, such as those needed for gravitational
wave physics. Secondly, extending the remit of the double copy
broadens our conceptual understanding as to whether this is a deep and
fundamental connection between different types of field theory, or
merely a coincidence for certain observables. If the former is true,
it suggests that our traditional way of formulating field theories may
be incomplete and / or hiding crucial underlying features.

If the double copy is a complete relationship between gauge and
gravity theories, it must somehow be applicable to all possible types
of solution. All previous examples involving amplitudes or classical
solutions (exact or otherwise) involve positive powers of the coupling
constants in the relevant theories. It remains to be seen whether or
not the double copy can be made truly non-perturbative,
i.e. applicable to strong-coupling solutions, containing negative
powers of the coupling. As a first step, a recent series of papers has
derived such solutions for biadjoint scalar field theory, with the
hope of matching them with known strong-coupling solutions in gauge
theory~\cite{White:2016jzc,DeSmet:2017rve,Bahjat-Abbas:2018vgo}. One
of these represents a monopole-like object, and
ref.~\cite{White:2016jzc} speculated that this might be relatable to
the well-known Wu-Yang monopole in gauge
theory~\cite{Wu:1967vp}. However, it has remained unclear how to
systematically construct such a double copy, whose rules must in any
case be fundamentally different to any previous case.\\

In this paper, we investigate the question of whether the biadjoint
monopole of ref.~\cite{White:2016jzc} maps to the Wu-Yang monopole,
and conclude that it does not. We present two arguments, where the
first involves infinitely boosting classical solutions in the various
theories we consider, and constructing shockwave solutions from
them. In gravity, this was first considered by Aichelburg and
Sexl~\cite{Aichelburg:1970dh}, who found that certain parameters have
to be rescaled when boosting, in order to keep physically measurable
effects finite. We will follow convention by referring to this general
procedure as an {\it ultraboost} in what follows. Related examples
including gravity and / or gauge theory can be found in
refs.~\cite{Lousto:1988ua,Lousto:1988ej,Lousto:1989ha,Lousto:1992th,Kozyulin:2011hk,Argurio:2008nb}. Certain
shockwave solutions are known to double
copy~\cite{Siegel:1999ew,Monteiro:2014cda,Saotome:2012vy}, and thus
comparing the ultraboosted biadjoint and Wu-Yang monopole solutions
allows us to confirm or refute whether or not they are connected by
the double copy. Indeed, we will see that, whilst the Wu-Yang monopole
survives its ultraboost, the biadjoint monopole does not, which seems
to indicate that they are not after all related. This conclusion is
not watertight, however, given that the physics in different theories
can turn out to be very different, even if the double copy relates
them.\\

In the second part of our study, we thus seek to explain why we could
have expected {\it a priori} that the biadjoint and Wu-Yang monopole
solutions are not related via the double copy. We recall the existence
of a singular gauge transformation that can be used to transform the
Wu-Yang solution into a non-abelian version of the Dirac magnetic
monopole in electromagnetism, whose form linearises the Yang-Mills
equations~\cite{Brandt:1979kk,Brandt:1980em}. The Dirac monopole is
known to double copy to the pure NUT solution in gravity. Furthermore,
its counterpart in biadjoint theory is already
known~\cite{Luna:2015paa}, and does not coincide with the
non-perturbative biadjoint monopole of
ref.~\cite{White:2016jzc}. Thus, there is no room for the
non-perturbative biadjoint monopole in matching up shockwaves, which
is consistent with it not surviving in the ultraboost limit. \\

Despite the negative result of our investigation, it proves to be
worthwhile for several reasons. Firstly, although both static
solutions and shockwaves are known to double-copy, the latter (in
gauge and biadjoint theory) have not been explicitly obtained from the
former using an ultraboost procedure in this context. The details of
how to perform an ultraboost in biadjoint theory are new, and the
comparison of this procedure with its gauge and gravity counterparts
proves interesting.  Secondly, the above-mentioned identification of
the Wu-Yang and Dirac monopoles indicates an emerging picture in which
both abelian- and non-abelian-like objects can double copy to the {\it
  same} gravity solution, which in turn suggests that the classical
double copy of ref.~\cite{Monteiro:2014cda} (which always concerns
abelian-like objects in the gauge theory) is more general than
previously thought. Such behaviour has been seen before in the study
of amplitudes~\cite{Oxburgh:2012zr}, and is consistent with the fact
that colour information is removed when taking the double copy.\\

The structure of our paper is as follows. In section~\ref{sec:review},
we review salient details concerning the classical double copy
procedure of ref.~\cite{Monteiro:2014cda}. In
section~\ref{sec:ultraboost}, we show how to obtain shockwaves which
have previously been shown to double copy via an ultraboost procedure
in biadjoint, gauge and gravity theories. We also find and interpret
the ultraboost of the non-perturbative biadjoint monopole of
ref.~\cite{White:2016jzc}. In section~\ref{sec:Dirac}, we review a
known singular gauge transformation relating the Wu-Yang and Dirac
monopoles, and explain the implications of this relationship for the
double copy. Finally, in section~\ref{sec:discuss}, we discuss our
results and conclude.

\section{The Kerr-Schild double copy}
\label{sec:review}

\begin{figure}
\begin{center}
\scalebox{0.8}{\includegraphics{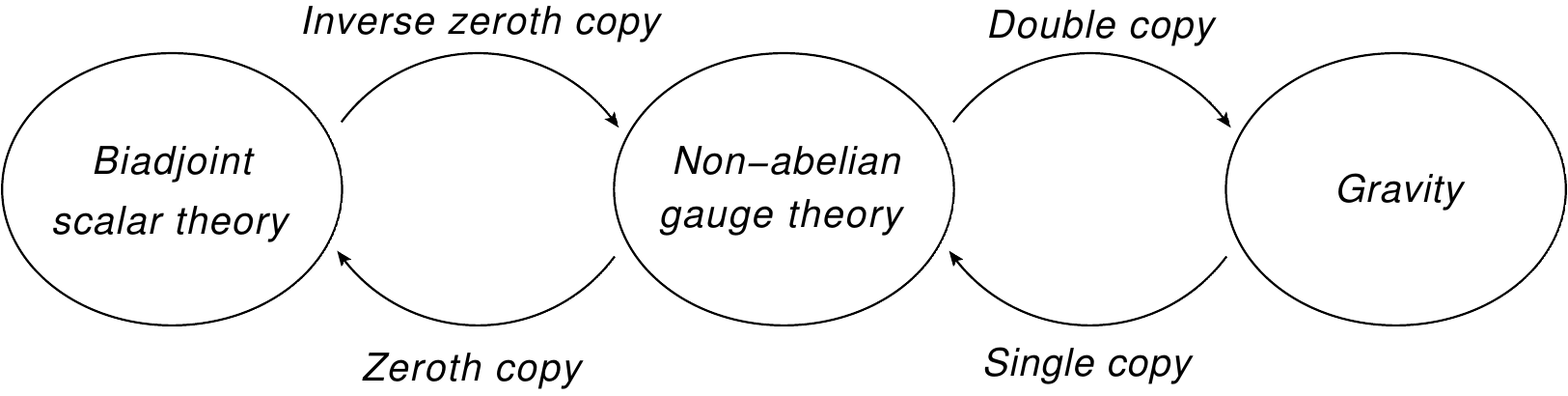}}
\caption{The different theories related by the double, single and
  zeroth copies.}
\label{fig:theories}
\end{center}
\end{figure}

The classical double copy of ref.~\cite{Monteiro:2014cda} is a
systematic procedure for relating certain exact classical solutions in
biadjoint, gauge and gravity theories. In the latter, we may define a
graviton field $h_{\mu\nu}$ via
\begin{equation}
g_{\mu\nu}=\eta_{\mu\nu}-\kappa h_{\mu\nu},\kappa=\sqrt{16\pi G_N},
\label{hdef}
\end{equation}
with $g_{\mu\nu}$ ($\eta_{\mu\nu}$) the full (Minkowski) metric with signature $(+,-,-,-)$,
and $G_N$ the Newton constant. We may then consider the family of {\it
  Kerr-Schild} solutions, for which the graviton has the special form
\begin{equation}
h_{\mu\nu}=\frac{\kappa}{2}\phi \,k_\mu k_\nu,
\label{hKS}
\end{equation}
where $\phi$ is a harmonic function, and the Kerr-Schild vector
$k^\mu$ is geodesic, and null with respect to either the Minkowski or
full metrics:
\begin{equation}
k\cdot \partial k^\mu=0,\quad g_{\mu\nu}k^\mu k^\nu=
\eta_{\mu\nu}k^\mu k^\nu=0.
\label{kconditions}
\end{equation}
Upon substituting the ansatz of eq.~(\ref{hKS}) into the Einstein
equations, they linearise, such that eq.~(\ref{hKS}) represents an
exact solution that is particularly tractable.  Given the quantities
$\phi$ and $k^\mu$ appearing in eq.~(\ref{hKS}), one may construct a
non-abelian gauge field
\begin{equation}
{\bf A}_\mu\equiv A_\mu^a {\bf T}^a,\quad A_\mu^a=c^a \phi k_\mu,
\label{AKS}
\end{equation}
where ${\bf T}^a$ is a generator of the gauge group with adjoint index
$a$, and $c^a$ an arbitrary colour
vector. Reference~\cite{Monteiro:2014cda} proved that for any
time-independent Kerr-Schild solution, the gauge field of
eq.~(\ref{AKS}) satisifes the linearised Yang-Mills equations, and
thus represents an exact physical solution of a non-abelian gauge
theory. Note that the procedure for obtaining $A_\mu^a$ is simply to
replace one copy of the Kerr-Schild vector $k_\mu$ with the colour
vector $c^a$ (correspondingly, a spacetime index of the field
$h_{\mu\nu}$ is replaced with a colour index in the field
$A_\mu^a$). One may repeat this procedure, so as to obtain a field
\begin{equation}
{\bf \Phi}=\Phi^{aa'}{\bf T}^a\, \tilde{\bf T}^{a'},\quad
\Phi^{aa'}=c^a\tilde{c}^{a'}\phi,
\label{PhiKS}
\end{equation}
where $\tilde{c}^{a'}$ is a second colour vector, that is potentially
associated with a different gauge group to that of $c^a$. As shown in
ref.~\cite{Monteiro:2014cda}, eq.~(\ref{PhiKS}) is a solution of a
linearisation of the biadjoint scalar field equation
\begin{equation}
\partial^2\Phi^{aa'}-\lambda f^{abc}\tilde{f}^{a'b'c'}\Phi^{bb'}\Phi^{cc'}=0,
\label{biadjoint}
\end{equation}
where $\lambda$ is a coupling constant, and $\{f^{abc}\}$,
$\{\tilde{f}^{a'b'c'}\}$ are sets of structure constants associated
with the two Lie groups. Starting with a time-independent Kerr-Schild
solution, we thus obtain solutions in a ladder of different field
theories, as depicted in figure~\ref{fig:theories}, where for
convenience we also display the name of the map between any given two
types of theory~\footnote{Figure~\ref{fig:theories} is not the whole
  story, but forms a subset of an ever-increasing web of theories
  related by double-copy-like transformations. See
  e.g. refs.~\cite{Carrasco:2015iwa,Cheung:2017ems,Bern:2019prr,Carrasco:2019qwr}
  for further details.}.\\

The above situation mirrors the existence of similar double copy
relations between amplitudes in these theories, such that our current
understanding is that the classical and amplitude double copies are
manifestations of the same underlying correspondence, and indeed
overlap where
relevant~\cite{Monteiro:2014cda,Luna:2015paa,Luna:2016due,Bahjat-Abbas:2017htu,Berman:2018hwd}. Despite
this, a number of open questions exist regarding the classical double
copy. Firstly, there is the issue of whether arbitrary sources, as
well as the fields themselves, can be furnished with a double-copy
interpretation~\cite{Ridgway:2015fdl,Luna:2016due,Carrillo-Gonzalez:2017iyj,Bah:2019sda}. This
is an interesting issue, which we will not explore in this
paper. Secondly, the family of time-independent Kerr-Schild metrics is
clearly very special, and a number of studies have tried to go beyond
this. Regarding time dependence, ref.~\cite{Monteiro:2014cda} pointed
out that certain infinite boosted (shockwave)
solutions~\cite{Coleman:1977ps,Aichelburg:1970dh}, whose
correspondence in gauge and gravity theories was first noted in
ref.~\cite{Siegel:1999ew}, can also be phrased in terms of the
Kerr-Schild double copy. The case of an arbitrarily accelerating point
particle was considered in ref.~\cite{Luna:2016due}, and related to
known scattering amplitudes for gluons and gravitons in the
Bremsstrahlung limit. Generalisation from time translation invariance
to an arbitrary Killing vector was considered in
ref.~\cite{Carrillo-Gonzalez:2017iyj}. There have also been attempts
to move beyond the simple Kerr-Schild
form. Reference~\cite{Luna:2015paa} considered the Taub-NUT solution
in gravity, which has a double Kerr-Schild form, in which the graviton
contains two terms of the form of eq.~(\ref{hKS}), involving different
harmonic functions and Kerr-Schild vectors. The latter obey certain
mutual orthogonality conditions, but this is no longer sufficient to
linearise the Einstein equations in general. Remarkably, linearisation
still occurs for the special case of Taub-NUT~\cite{Chong:2004hw}, and
the single copy is found to be a point-like dyon, carrying both
electric and magnetic charge~\footnote{Recently, the known {\it
    electromagnetic duality} relating electric and magnetic charges
  has been explored from a double copy point of
  view~\cite{Huang:2019cja,Alawadhi:2019urr,Banerjee:2019saj}.}. More
generally, one may consider the classical double copy of
non-Kerr-Schild solutions, but at the price of having to work
order-by-order in perturbation
theory~\cite{Luna:2016hge,Luna:2017dtq,Goldberger:2016iau,Goldberger:2017frp,Goldberger:2017ogt}. Thus,
Kerr-Schild coordinates are not a fundamental prequisite for being
able to construct a classical double copy. Rather, they constitute a
maximally convenient case, in which it is possible to make statements
to all orders in perturbation theory~\footnote{A related programme of
  work has shown that a classical double copy is possible in arbitrary
  coordinate systems, if one restricts to linearised level
  only~\cite{Anastasiou:2014qba,Borsten:2015pla,Anastasiou:2016csv,Anastasiou:2017nsz,Anastasiou:2018rdx,Cardoso:2016ngt,LopesCardoso:2018xes,Borsten:2019prq}.}. \\

A third puzzle regarding the Kerr-Schild double copy concerns the fact
that the single copy of the gravity solution is always abelian-like,
in that it linearises the Yang-Mills equations (n.b. the gauge theory
solution is still formally non-abelian, in that it is dressed by the
colour dependence of eq.~(\ref{AKS})). Whilst this follows from the
mathematical arguments of ref.~\cite{Monteiro:2014cda}, this situation
is slightly at odds with the double copy story for amplitudes, in
which the non-abelian nature of the gauge theory plays a pivotal role,
through the so-called {\it BCJ duality}~\cite{Bern:2008qj} relating
colour and kinematic information. It would perhaps be more desirable
if one could associate fully non-linear solutions in the gauge theory
with gravitational counterparts, and there is also the possibility
that both abelian-like and fully non-linear solutions in the gauge
theory may map to the {\it same} gravity solution. There is in fact a
precedent for this in an amplitudes context~\cite{Oxburgh:2012zr}, and
we will return to this in what follows.\\

Before moving on, it is worth providing those examples of the
Kerr-Schild double copy that we will later rely on. Arguably the
simplest example is that of a pointlike mass sourcing a Schwarzschild
black hole, for which the quantities appearing in eq.~(\ref{hKS}) are
given by 
\begin{equation}
\phi=\frac{M}{4\pi r},\quad k_\mu=\left(1,\frac{\vec{x}}{r}\right),
\label{Schwarzschild}
\end{equation}
where $\vec{x}=(x,y,z)$ is the radial position vector in Cartesian
coordinates. The single copy of this is a point charge at the origin,
given by
\begin{equation}
A^a_\mu=\frac{g c^a}{4\pi r} k_\mu,
\label{SchwarzschildA}  
\end{equation}
and the zeroth copy is then
\begin{equation}
\Phi^{aa'}=\frac{\lambda c^a \tilde{c}^{a'}}{4\pi r}.
\label{SchwarzschildPhi}  
\end{equation}
Another case we will utilise is that of the {\it Aichelburg-Sexl
  solution} in gravity, representing a shockwave moving along the $x$
direction at the speed of light, and such that the gravitational
impulse imparted to a stationary test particle is
finite~\cite{Aichelburg:1970dh}. This
has a Kerr-Schild form, with:
\begin{equation}
\phi=-\frac{M}{4\pi}\log\left(\frac{\rho}{\rho_0}\right)\delta(u),\quad
k_\mu=(1,-1,0,0),
\label{ASKS}
\end{equation}
where 
\begin{equation}
u=t-x
\label{udef}
\end{equation}
is a light-cone coordinate, 
\begin{equation}
\rho=\sqrt{y^2+z^2}
\label{rhodef}
\end{equation}
the cylindrical radius as measured from the $x$ axis, $\rho_0$ an
arbitrary constant, and the vector $k_\mu$ is again expressed in
Cartesian coordinates. The single copy of this solution is a
non-abelian plane wave~\cite{Coleman:1977ps}
\begin{equation}
A_\mu^a=-\frac{g c^a}{4\pi}
\log\left(\frac{\rho}{\rho_0}\right)\delta(u)k_\mu,
\label{Aplanewave}
\end{equation}
and there is also a corresponding biadjoint zeroth copy
\begin{equation}
\Phi^{aa'}=-\frac{\lambda c^a\tilde{c}^{a'}}{4 \pi}
\log\left(\frac{\rho}{\rho_0}\right)\delta(u).
\label{Phiplanewave}
\end{equation}
In each theory, the factor $\delta(u)$ confines the influence of the
field to a plane transverse to the $x$-direction, moving at light
speed. There is a non-trival profile function, which depends only on
the transverse coordinates $(y,z)$~\footnote{The above shockwave
  solutions provide a nice link between the classical and amplitude
  double copies: the gauge and gravity shockwaves were constructed
  from an all-order Feynman diagram analysis in
  ref.~\cite{Saotome:2012vy}. }. \\

All of the above examples involve positive powers of the coupling
constants in the biadjoint, gauge or gravity theories. It remains
unknown whether or not the double copy can be extended to
non-perturbative solutions, involving inverse powers of the
coupling. Solutions of the biadjoint theory play a crucial role in
both the amplitude and Kerr-Schild double
copies~\cite{Monteiro:2014cda}. Thus, it seems natural to assume that
this should also be the case in a nonperturbative correspondence,
should the latter exist. To this end,
refs.~\cite{White:2016jzc,DeSmet:2017rve,Bahjat-Abbas:2018vgo}
initiated the programme of cataloguing non-linear solutions of
biadjoint theory. The simplest such solution has the form of a static
spherically symmetric monopole-like object residing at the origin of
spacetime:
\begin{equation}
\Phi^{aa'}=-\frac{2\delta^{aa'}}{\lambda T_A r^2},
\label{monopole}
\end{equation}
where it is assumed that both Lie groups in the biadjoint theory are
the same, and the constant $T_A$ is defined in terms of the structure
constants via
\begin{equation}
f^{abc}f^{a'bc}=\delta^{aa'} T_A.
\label{TAdef}
\end{equation}
For the specific case in which the common gauge group is SU(2), there
is also a continuous family of solutions given by
\begin{equation}
\Phi^{aa'}=\frac{1}{\lambda r^2}\left[-k\left(\delta^{aa'}-\frac{x^a x^{a'}}
{r^2}\right)\pm\sqrt{2k-k^2}\frac{\epsilon^{aa'd}x^d}{r}\right],\quad
0\leq k\leq 2.
\label{monopole2}
\end{equation}
There was already some speculation in ref.~\cite{White:2016jzc} about
whether any of these solutions could be related to pointlike objects
in non-abelian gauge theory. In the case of SU(2), a natural candidate
is the {\it Wu-Yang monopole} of ref.~\cite{Wu:1967vp}, which in a
particular gauge takes the form
\begin{equation}
A_0^a=0,\quad A_i^a=-\frac{\epsilon_{iak}x^k}{g r^2},
\label{WuYang}
\end{equation}
where $g$ is the coupling constant. Like the solutions of
eq.~(\ref{monopole}, \ref{monopole2}), this contains an inverse power
of the coupling, and also has a pure power-like dependence in the
spherical radius $r$, where the power itself can be dictated on
dimensional grounds~\footnote{Note that $g$ in the gauge theory is
  dimensionless in four spacetime dimensions, whereas the coupling
  constant $\lambda$ in eq.~(\ref{biadjoint}) has dimensions of
  mass.}. Given the lack of any precedent for how to formulate a
non-perturbative double copy, ref.~\cite{Monteiro:2014cda} left as
merely speculative the suggestion that the biadjoint monopoles of
eqs.~(\ref{monopole}, \ref{monopole2}) are related to the Wu-Yang
monopole. Our aim here is to examine this systematically, and the
starting point will be the above-mentioned fact that shockwave
solutions are known to double copy. By ultraboosting the biadjoint
monopole, it may turn out to have properties corresponding to a known
shockwave, or exhibit other simplifying features that enable a
suitable double copy interpretation to be obtained. If it instead does
not survive the ultraboost, then this is evidence that the speculative
link between the biadjoint and Wu-Yang monopoles may in fact be
incorrect. 

\section{Shockwaves and ultraboosts}
\label{sec:ultraboost}

Above, we have suggested ultraboosting monopole solutions in biadjoint
and gauge theories, in order to see what can be learned about the
possible existence of a nonperturbative double copy. Before
considering the case of the biadjoint monopole, however, it pays to
revisit the Aichelburg-Sexl family of shockwaves, written here in
eqs.~(\ref{ASKS}, \ref{Aplanewave}, \ref{Phiplanewave}). We will
recast the ultraboost procedure of ref.~\cite{Aichelburg:1970dh} using
Kerr-Schild coordinates, so that double copy properties are
manifest. This will also allow us to examine the ultraboost in the
biadjoint scalar theory, which has not been previously considered.\\

\subsection{The Aichelburg-Sexl shockwave in Kerr-Schild coordinates}
\label{sec:ASKS}

Let us begin with the point mass (Schwarzschild) solution of
eqs.~(\ref{hKS}, \ref{Schwarzschild}), taken to be stationary in an
inertial frame $S'$ with Cartesian coordinates $(t',x',y',z')$. We may
then consider that $S'$ is moving with boost parameter $\beta\equiv v$
(in natural units) in the $+x$ direction relative to a second frame
$S$ whose coordinates are $(t,x,y,z)$. The two sets of coordinates are
related by the Lorentz transformation
\begin{equation}
\left(\begin{array}{c}t'\\x'\\y'\\z'\end{array}\right)=
\left(\begin{array}{cccc}\gamma & -\gamma\beta & 0 & 0\\
-\gamma\beta & \gamma & 0 & 0 \\
0 & 0 & 1 & 0\\
0 & 0 & 0 & 1\end{array}\right)\left(\begin{array}{c}
t \\ x \\ y \\ z\end{array}\right)=
\left(\begin{array}{c}
\gamma(t-\beta x) \\ \gamma(x-\beta t) \\ y \\ z\end{array}\right)
.
\label{Lorentz}
\end{equation}
The graviton in $S'$ is given in Kerr-Schild form by 
\begin{displaymath}
h'_{\mu\nu}=\frac{\kappa}{2}\phi(x')k'_\mu k'_\nu,
\end{displaymath}
where the function $\phi$ and Kerr-Schild vector are given in terms of
the primed coordinates by eq.~(\ref{Schwarzschild}). Boosting these
ingredients to the unprimed frame, one finds
\begin{align}
\phi(x)&=\frac{M}{4\pi}\frac{1}{[\gamma^2(x-\beta
    t)^2+\rho^2]^{1/2}},\notag\\
k_\mu&=\left(\gamma-\frac{\gamma^2\beta(x-\beta t)} {[\gamma^2(x-\beta
    t)^2+\rho^2]^{1/2}}, -\gamma\beta+\frac{\gamma^2(x-\beta
  t)}{[\gamma^2(x-\beta t)^2+\rho^2]^{1/2}},
\frac{y}{[\gamma^2(x-\beta t)^2+\rho^2]^{1/2}},\right.\notag\\
&\left.\quad\quad
\frac{z}{[\gamma^2(x-\beta t)^2+\rho^2]^{1/2}}\right).
\label{phikboost}
\end{align}
To look for a shockwave solution, we must take the limit
$\gamma\rightarrow \infty$, whilst also regularising the solution so that
physically measurable quantities are finite. One such quantity is the
deflection of a test particle upon crossing the shockwave, which is
linear in the field. Thus, we require that $h_{\mu\nu}$ is finite in
the ultraboost limit. However, the limiting procedure itself is rather
subtle, given that one finds different results for the limiting values
of the quantities in eq.~(\ref{phikboost}) depending upon whether one
is inside ($x=\beta t$) or outside ($x\neq \beta t$) the plane of the
shockwave. For the former case one obtains
\begin{align}
\phi\xrightarrow{\gamma\rightarrow\infty}
\frac{M}{4\pi}\frac{1}{\rho}+{\cal O}(\gamma^{-1}),\quad
k_\mu \xrightarrow{\gamma\rightarrow\infty}\gamma \bar{k}_\mu
+{\cal O}(\gamma^0),
\quad x=\beta t,
\label{phikboost2}
\end{align}
where we have defined the dimensionless 4-vector
\begin{equation}
\bar{k}_\mu=(1,-1,0,0).
\label{barkmu}
\end{equation}
Outside the shockwave plane one obtains
\begin{align}
\phi\xrightarrow{\gamma\rightarrow\infty}
\frac{M}{4\pi}\frac{1}{\gamma|t-x|}+{\cal O}(\gamma^{-2}),\quad
k_\mu \xrightarrow{\gamma\rightarrow\infty}2\gamma \theta(t-x)
\bar{k}_\mu
+{\cal O}(\gamma^0),
\quad x\neq\beta t,
\label{phikboost3}
\end{align}
where $\theta(t-x)$ is the Heaviside function. The complete boosted
graviton field is then given by
\begin{align}
h_{\mu\nu}\xrightarrow{\gamma\rightarrow\infty}
\frac{\kappa}{2}\frac{M}{4\pi}\begin{cases}
\frac{\gamma^2}{\rho}\bar{k}_\mu\bar{k}_\nu+{\cal O}(\gamma),
\quad x=\beta t\\
\frac{4\gamma}{|t-x|}\theta(t-x)\bar{k}_\mu\bar{k}_\nu+{\cal O}(\gamma^0),
\quad x\neq \beta t.
\end{cases}
\label{hboost}
\end{align}
It is clear that things are badly divergent, both on and off the
plane, which makes physical sense: boosting a massive particle to
light speed requires infinite energy, and thus will result in a
divergent field configuration! The transition to a finite shockwave
solution proceeds as follows~\cite{Aichelburg:1970dh}. First, one
may rescale the mass\footnote{In the literature this rescaling is noted to keep the energy finite whilst taking the rest mass to zero. Physically one may view this as a necessary step in changing description between a massive and a massless particle.} according to
\begin{equation}
M\rightarrow \frac{M}{\gamma}.
\label{mrescale}
\end{equation}
Then the graviton field of eq.~(\ref{hboost}) remains infinite inside
the plane $x=\beta t$, but not outside it. This suggests that the
ultraboosted field could indeed contain a delta function $\delta(u)$,
localising the extent of the field to the shockwave plane only. To
recognise the delta function, we may reinstate the $\gamma$-dependence
for the field near the shockwave plane, writing this as
\begin{equation}
h_{\mu\nu}\xrightarrow{\gamma\rightarrow \infty, M\rightarrow M/\gamma}
\frac{\kappa}{2}\frac{\gamma M}{4\pi}\frac{1}{[\gamma^2(x-\beta t)^2+\rho^2]^{1/2}} \,
\bar{k}_\mu \bar{k}_\nu.
\label{hboost2}
\end{equation}
We may then attempt to use the general formula (see
e.g. ref.~\cite{Argurio:2008nb})
\begin{equation}
\lim_{\gamma\rightarrow\infty}\gamma f(\gamma u)=\delta(u)
\int_{-\infty}^\infty dw\,f(w)
\label{deltalim}
\end{equation}
for expressing a Dirac delta function as the limit of a {\it delta
  sequence} $f(u)$, namely a function that can be continuously
transformed to make an infinitely sharp peak at $u=0$. In the present
case, we may identify  
\begin{equation}
f(u)=\frac{M}{4\pi}\frac{1}{(u^2+\rho^2)^{1/2}},
\label{fudef}
\end{equation}
such that eq.~(\ref{deltalim}) indeed corresponds to taking the limit
of the prefactor in eq.~(\ref{hboost2}). However, application of
eq.~(\ref{deltalim}) then fails due to the fact that the integral on
the right-hand side is not convergent. We will interpret the physics
of this shortly, but for now note that we can modify the function
$f(u)$ according to~\footnote{Reference~\cite{Aichelburg:1970dh}
  justifies this transformation based on the fact that it corresponds
  to a diffeomorphism of the graviton field.}
\begin{equation}
f(u)\rightarrow \frac{M}{4\pi}\left[\frac{1}{(u^2+\rho^2)^{1/2}}
-\frac{1}{(u^2+\rho_0^2)^{1/2}}\right],
\label{fmodify}
\end{equation}
such that the integral on the right-hand side of eq.~(\ref{deltalim})
becomes 
\begin{align}
\int_{-\infty}^\infty du\,f(u)&=
\frac{M}{4\pi}\lim_{\epsilon\rightarrow 0}\int_{-\infty}^\infty du
\left[(u^2+\rho^2)^{-1/2+\epsilon}-(u^2+\rho_0^2)^{-1/2+\epsilon}
\right]\notag\\
&=\frac{M}{4\pi}\lim_{\epsilon\rightarrow 0}
\left\{\frac{\sqrt{\pi}\,\Gamma(-\epsilon)}{\Gamma(1/2-\epsilon)}
\left[(\rho^2)^{\epsilon}-(\rho_0^2)^{\epsilon}\right]\right\},
\label{fint}
\end{align}
where we have chosen to introduce a regularisation parameter
$\epsilon$. The latter reveals that each of the separate terms in
eq.~(\ref{fint}) is divergent, but that the combination produces the
well-defined limit
\begin{equation}
\int_{-\infty}^\infty du\,f(u)=-\frac{M}{4\pi}\log\left(\frac{\rho^2}
{\rho_0^2}\right).
\label{flim}
\end{equation}
The complete form for the ultraboosted field is now
\begin{equation}
h_{\mu\nu}\xrightarrow{\gamma\rightarrow \infty, \, M\rightarrow M/\gamma}
-\frac{\kappa}{2}\frac{M}{4\pi}\log\left(\frac{\rho}{\rho_0}\right)\delta(u)\bar{k}_\mu
\bar{k}_\nu,
\label{hboostres}
\end{equation}
which is precisely the Aichelburg-Sexl shockwave of eq.~(\ref{ASKS}). We can
now interpret the divergence that appeared in trying to apply
eq.~(\ref{deltalim}) to the original function of eq.~(\ref{fudef}):
its regularisation merely amounts to introducing an overall constant,
that sets the scale of the logarithmic prefactor in
eq.~(\ref{hboostres}), and which has to be present on dimensional
grounds. It has no physical consequences, given that deflections of
test particles depend only upon derivatives of the metric.\\

Note that in Kerr-Schild coordinates, the combination of the mass
rescaling and regularisation procedures can be rephrased in a
particularly simple form. The correct ultraboost of the Schwarzschild
solution to form an Aichelburg-Sexl shockwave consists of the
following two steps:
\begin{enumerate}
\item[(i)] One must rescale the mass parameter according to
  eq.~(\ref{mrescale}).
\item[(ii)] One must modify the boosted Kerr-Schild function in
  eq.~(\ref{phikboost}) according to
\begin{equation}
\phi(x)\rightarrow \frac{M}{4\pi}
\left[\frac{1}{[\gamma^2(x-\beta t)^2+\rho^2]^{1/2}}
-\frac{1}{[\gamma^2(x-\beta t)^2+\rho_0^2]^{1/2}}
\right].
\label{phimodify}
\end{equation}
\end{enumerate}
This is slightly different to the above analysis, in that one modifies
the $\phi$ function before taking the limit
$\gamma\rightarrow\infty$. Importantly, this modification maintains
the fact that $\phi$ is harmonic, so that the Kerr-Schild conditions
still apply. Repeating the above analysis yields eq.~(\ref{hboostres})
as before~\footnote{An advantage of this alternative way of
  formulating the ultraboost is that the boosted field manifestly
  vanishes off of the shockwave plane, in contrast to
  eq.~(\ref{hboost}). Indeed, the latter behaviour is associated with
  the fact that the original $\phi$ function could not be used to form
  a delta sequence.}. By modifying $\phi$ itself, however, we have
cast the ultraboost procedure into a form which has a natural
counterpart under the single and zeroth copies. 

\subsection{Single and zeroth copies}
\label{sec:ASgauge}

In the previous section, we have given an explicit procedure for
ultraboosting a point mass in Kerr-Schild coordinates. We may carry
out a similar procedure for the point charge in a gauge theory, for
which one may start with the single copy of the Schwarzschild solution
in the primed coordinate system, and boost it to obtain the following
gauge field:
\begin{equation}
{\bf A}_\mu=\phi\,k_\mu.
\label{Amuboosted}
\end{equation}
Here $\phi$ is given by the function of eq.~(\ref{phikboost}), but
with the mass replaced by the colour charge $c^a {\bf T}^a$, and
$k_\mu$ is also given in eq.~(\ref{phikboost}). For the ultraboost, we
may again modify $\phi$ according to eq.~(\ref{phimodify}), after
which we find the following limit:
\begin{equation}
{\bf A}_\mu\xrightarrow{\gamma\rightarrow\infty}
-\frac{c^a{\bf T}^a}{4\pi}\log\left(\frac{\rho}{\rho_0}\right)
\delta(u)\bar{k}_\mu,
\label{Aboostedres}
\end{equation}
where $\bar{k}_\mu$ has been defined in eq.~(\ref{barkmu}). This is
precisely the form of the gauge theory shockwaves considered in
refs.~\cite{Coleman:1977ps,Saotome:2012vy} (see also
ref.~\cite{Kozyulin:2011hk} for an interesting discussion of gauge
theory shockwaves in a different context). Note that to obtain a
finite field configuration, there is no need to rescale the charge, as
was needed for the mass parameter in the gravity theory. We will
return to this point below. \\

For the zeroth copy, we may simply consider the biadjoint field (in
the primed coordinate system)
\begin{equation}
\Phi^{aa'}=\frac{\lambda c^a\tilde{c}^{a'}}{4\pi r'}.
\label{Phiprimed}
\end{equation}
Boosting to the unprimed system and making the modification of
eq.~(\ref{phimodify}), one obtains the limit
\begin{equation}
\Phi^{aa'}\xrightarrow{\gamma\rightarrow\infty, \, c^a\tilde{c}^{a'}
\rightarrow \gamma c^a\tilde{c}^{a'}}
-\frac{\lambda c^a\tilde{c}^{a'}}{4\pi}
\log\left(\frac{\rho}{\rho_0}\right)
\delta(u),
\label{Phiboosted}
\end{equation}
where in order to achieve a non-vanishing field configuration, we must
rescale the charges as shown. Upon doing so, the shockwave solutions
of eqs.~(\ref{hboostres}, \ref{Aboostedres}, \ref{Phiboosted}) are
related directly by the usual single and zeroth copy procedures. That
is, one may write eq.~(\ref{hboostres}) in the form
\begin{equation}
h_{\mu\nu}=\frac{\kappa}{2}\bar{\phi}\bar{k}_\mu\bar{k}_\nu,\quad
\bar{\phi}=-\frac{M}{4\pi}\log\left(\frac{\rho}{\rho_0}\right)\delta(u),
\label{hboostedKS}
\end{equation}
where $\bar{k}_\mu$ (from eq.~(\ref{barkmu})) is indeed null and
geodesic, and $\bar{\phi}$ harmonic. The single and zeroth copies
imply that one must remove factors of $\bar{k}_\mu$, and replace mass
with charge accordingly, leading directly to eqs.~(\ref{Aboostedres})
and~(\ref{Phiboosted}). We can also make sense, from a double copy
point of view, of the modification of eq.~(\ref{phimodify}). As
stressed in ref.~\cite{Monteiro:2014cda}, the function $\phi$ in
eq.~(\ref{hKS}) can be interpreted as a scalar propagator, and is
analogous to the denominator factors in amplitudes, which are not
modified upon taking the double copy from gauge theory to gravity. In
the present case, the function $\bar{\phi}$ that we arrive at after
the ultraboost is indeed the known propagator in two spatial
dimensions (i.e. corresponding to the transverse plane), and the
modification of eq.~(\ref{phimodify}) is necessary so as to construct
the most general form of the propagator by including the constant
$\rho_0$. This in turn explains why eq.~(\ref{phimodify}) is necessary
when performing ultraboosts in all three theories of
figure~\ref{fig:theories}.\\

Above, we have seen that different scalings of mass / charge
parameters in different theories are necessary, to obtain a finite
field. The sources for the biadjoint, gauge and gravity theories
respectively are as follows:
\begin{equation}
\rho^{aa'}=c^a\tilde{c}^{a'}\delta^{(3)}(\vec{x}),\quad
j_\mu^{a}=c^a\delta_\mu^{0}\delta^{(3)}(\vec{x}),\quad
T_{\mu\nu}=M\delta_\mu^{0}\delta_\nu^{0}\,\delta^{(3)}(\vec{x}),
\label{sources}
\end{equation}
The rescalings act on the parameters entering these source terms, and
we summarise the different findings in table~\ref{tab:rescalings}.
\begin{table}
\begin{center}
\begin{tabular}{c|c|c}
Theory & Parameter & Rescaling\\
\hline
Biadjoint scalar & $c^{a}\tilde{c}^{a'}$ & $\gamma$\\
Gauge theory & $c^a$ & 1\\
Gravity & $M$ & $\gamma^{-1}$\\
\end{tabular}
\caption{Different scalings needed for the source parameters in
  different theories, where the factor in the third column multiplies
  the parameter in question.}
\label{tab:rescalings}
\end{center}
\end{table}
Confirmation that these are indeed the correct scalings to obtain
finite physical effects can be verified by placing a test particle
away from the origin~\footnote{The shockwave profile diverges at
  $y=z=0$, so that we must place the test particle elsewhere.} in the
plane $x=0$, and calculating the impulse
\begin{equation}
\delta p_\mu=\int_{-\infty}^\infty dt\,\frac{d p_\mu}{dt}
\label{impulse}
\end{equation}
that it receives as the shockwave passes. In biadjoint theory the
integrand is given by~\cite{Goldberger:2017frp}
\begin{equation}
\frac{dp_\mu}{dt}=-\frac{\lambda}{m} c_2^a\tilde{c}_2^{a'}\partial_\mu \Phi^{aa'},
\label{biadjointforce}
\end{equation}
where $c_2^{a}$ and $\tilde{c}_2^{a'}$ are the colour charge vectors
associated with the test particle. Substituting eq.~(\ref{Phiboosted})
into eq.~(\ref{biadjointforce}) and integrating according to
eq.~(\ref{impulse}), one finds (see appendix \ref{app:impulse} for more details)
\begin{equation}
\delta p^{\mu}_{\rm scal.}
=-\frac{\lambda^2}{m}\frac{ c\cdot c_2\,\tilde{c}\cdot\tilde{c}_2}{2 \pi\rho}
(0,0,\hat{\bf \rho}),
\label{biadjointimpulse}
\end{equation}
where $\hat{\bf \rho}$ is a unit vector in the radial direction in the
plane $x=0$. This is indeed finite as required. Furthermore, it does
not depend on the constant $\rho_0$, justifying the remarks made
above. For the gauge theory, one may use the Lorentz force law
\begin{equation}
\frac{dp_\mu}{dt}=g c_2^a F^{a}_{\mu\nu}\, v^\nu,
\label{lorentzforce}
\end{equation}
where $F^a_{\mu\nu}$ is the field strength tensor of the shockwave,
$c_2^a$ the colour charge vector of the test particle, whose initial
velocity is $v_\mu=(1,\vec{0})$. One then finds an impulse
\begin{equation}
\delta p^{\mu}_{\rm gauge}=g^2\frac{c_2\cdot c}{2\pi \rho}(0,0,\hat{\bf \rho}).
\label{impulsegauge}
\end{equation}
Finally, for gravity one may use the geodesic equation
\begin{equation}
\frac{d p^\mu}{dt}=-m\Gamma^\mu_{\;\nu\sigma} v^\nu v^\sigma,
\label{geodesicequation}
\end{equation}
where $\Gamma^\mu_{\nu\lambda}$ is the Christoffel symbol. One
subsequently obtains the impulse
\begin{equation}
\delta p^{\mu}_{\rm grav.}=-\frac{\kappa^2 Mm}{8\pi \rho}
(0,0,\hat{\bf \rho}).
\label{impulsegrav}
\end{equation}
In all cases, the impulses are finite, which indeed is entirely
consistent with the fact that the rescalings of
table~\ref{tab:rescalings} were such as to make the profiles of the
fields in the shockwave plane finite: the three force laws of
eqs.~(\ref{biadjointforce}, \ref{lorentzforce},
\ref{geodesicequation}) are all linear in their respective
fields~\footnote{It is worth pointing out that different physical
  quantities may require different scalings upon performing an
  ultraboost, such as the non-linear electromagnetic interactions
  considered in ref.~\cite{Kozyulin:2011hk}.}. Turning this around, we
can use physical finiteness of the impulses to dictate the rescaling
needed in each theory, and the power of $\gamma$ needed to rescale the
parameters of the field can then simply be traced to the number of
factors of the the Kerr-Schild vector each field contains, and hence
its spin.\\

In this section, we have shown how the ultraboost procedure can be
implemented in biadjoint, gauge and gravity theories, so as to be
fully compliant with the double copy. We focused on point charges or
masses that are solutions of the linearised field equations in each
case (up to a source term localised at the origin). In all cases, the
charges survived the ultraboost, resulting in a well-defined shockwave
solution that could be meaningfully mapped between theories as in
figure~\ref{fig:theories}. Armed with this experience, let us now see
what happens if we try to ultraboost the non-perturbative biadjoint
monopole.

\subsection{Ultraboosting the biadjoint monopole}
\label{sec:biadointboost}

Starting with the solution of eq.~(\ref{monopole}) in the rest frame
$S'$, we may boost to the unprimed frame to obtain
\begin{equation}
\Phi^{aa'}=-\frac{2\delta^{aa'}}{\lambda T_A}\frac{1}{[\gamma^2(x-\beta
    t)^2+\rho^2]}.
\label{monopoleboost}
\end{equation}
Taking the ultraboost limit, one finds
\begin{equation}
\Phi^{aa'}\xrightarrow{\gamma\rightarrow \infty} 
-\frac{2\delta^{aa'}}{\lambda T_A}\begin{cases}
\frac{1}{\rho^2},\quad x=\beta t\\
{\cal O}(\gamma^{-2}),\quad x\neq \beta t.
\end{cases}
\label{monopoleboost2}
\end{equation}
Due to the different dependence on the radial coordinate, this does
not diverge on the transverse plane, in contrast to the boosted point
charge considered in the previous section. Without this divergence,
eq.~(\ref{monopoleboost2}) does not constitute a shockwave: it is at
most a tepid ripple. That is, it imparts no finite impulse to a test
particle, given that a finite field is confined to an infinitely thin
plane.\\

It is interesting to note that one could still obtain a shockwave by
rescaling the coupling $\lambda$ in eq.~(\ref{monopoleboost2}). Carrying out
a similar analysis to the previous section, one may show that
\begin{equation}
\Phi^{aa'}\xrightarrow{\gamma\rightarrow\infty, \, \lambda\rightarrow \gamma^{-1}\lambda}
-\frac{2\delta^{aa'}}{\lambda T_A}\frac{1}{\rho}\delta(u),
\label{phiboostrescale}
\end{equation}
which does indeed exert a finite impulse. However, the nature of this
rescaling is very different to that of the boosted point charges
considered earlier. In the latter cases, the parameters entering the
source terms (i.e. masses and charges) were rescaled, which does not
change the field equations themselves. Rescaling the coupling for the
non-perturbative monopole, however, does indeed constitute changing
the field equations, and hence the theory~\footnote{Coupling constants
  are of course not constant when quantum corrections are included,
  leading to renormalisation. But that is not what is happening
  here.}. \\

Based on the above considerations, we conservatively conclude that the
non-perturbative biadjoint monopole does not survive an
ultraboost. Similar considerations are reached for the more general
monopoles of eq.~(\ref{monopole2}), as we show in
appendix~\ref{app:generalmonopole}. There are then two possibilities
as regards a potential non-perturbative double copy, and in particular
the suggestion that the biadjoint monopole could be related to a
Wu-Yang monopole in gauge theory. The first is that the two objects
are indeed related, but that the physics of ultraboosting is
potentially very different in the two theories, such that the
biadjoint monopole disappears. Whether or not the Wu-Yang monopole
survives an ultraboost is irrelevant for the argument. The second
possibility is that the biadjoint monopole disappears because it does
not need to match up with a known shockwave solution, and in
particular is not related to the Wu-Yang monopole. In the next
section, we will explain why it is in fact this second possibility
that appears to be correct.

\section{Relating the Wu-Yang and Dirac monopoles}
\label{sec:Dirac}

Above, we saw that the failure of the biadjoint monopole to survive
its ultraboost is possible evidence for its not being related to the
Wu-Yang monopole after all, contrary to the speculation of
ref.~\cite{White:2016jzc}. We now explain why this must be the case,
and our explanation will itself provide new insights into the remit of
the classical double copy itself.\\

First, we recall that there is a gauge transformation that relates the
Wu-Yang monopole in SU(2) gauge theory to a non-abelian embedding of
the Dirac monopole, as noted in
refs.~\cite{Brandt:1979kk,Brandt:1980em}~\footnote{A similar gauge
  transformation is used to relate different forms of the 't
  Hooft-Polyakov monopole of
  refs.~\cite{'tHooft:1974qc,Polyakov:1974wq}. }. By the latter, we
mean a gauge field of the form
\begin{equation}
A^a_\mu=c^a A_\mu^{\rm Dirac},
\label{ADform} 
\end{equation}
where $c^a$ is a constant colour vector as usual, and 
\begin{equation}
A_\mu^{\rm Dirac}=\left(0,-\frac{\tilde{g}y}{r(r+z)},
\frac{\tilde{g}x}{r(r+z)},0\right)
\label{ADirac}
\end{equation}
is a solution of the Maxwell equations corresponding to a Dirac
(magnetic) monopole of charge $\tilde{g}$. This has a well-known
string singularity, which we have chosen to extend from the origin
along the $-z$ direction. Note that eq.~(\ref{ADirac}) is reminiscent
of the single copy solutions in the Kerr-Schild double copy, in that
it is manifestly of a form which linearises the Yang-Mills
equations. We may thus refer to this solution as {\it abelian-like},
even though it is not, strictly speaking, a solution of an abelian
gauge theory. \\

Without loss of generality for the following arguments, we may choose
the constant colour vector to lie in the $3$-direction in the internal
space, so that $c^a=\delta^{a3}$. Given that we are focusing on an
SU(2) gauge group, we may thus write the complete (matrix-valued)
gauge field as
\begin{equation}
{\bf A}_\mu=A_\mu^{\rm Dirac}{\bf \sigma}_3,
\label{Amatrix}
\end{equation}
where 
\begin{equation}
{\bf \sigma}_i=\frac{1}{2}{\bf \tau}_i,
\label{sigmadef}
\end{equation}
is a generator of SU(2), in terms of the Pauli matrices
\begin{equation}
{\bf \tau}_1=\left(\begin{array}{rr} 0 & 1 \\ 1 & 0
\end{array}\right),\quad
{\bf \tau}_2=\left(\begin{array}{rr} 0 & -i \\ i & 0
\end{array}\right),\quad
{\bf \tau}_3=\left(\begin{array}{rr} 1 & 0 \\ 0 & -1
\end{array}\right).\quad
\label{Pauli}
\end{equation}
We may transform eq.~(\ref{ADirac}) to spherical polar coordinates
$(t,r,\theta,\phi)$, yielding
\begin{equation}
{\bf A}_\mu={\bf \sigma}_3\left(0,0,0,\frac{\tilde{g}(1-\cos\theta)}{r\sin\theta}
\right)_{\rm spherical}.
\label{Amatrix2}
\end{equation}
Next, we can make a gauge transformation
\begin{equation}
{\bf A}_\mu\rightarrow {\bf A}'_\mu=
{\bf U}\,{\bf A}_\mu\,{\bf U}^{-1}+\frac{i}{g}{\bf U}
\left(\partial_\mu {\bf U}^{-1}\right).
\label{Atransform}
\end{equation}
Here $g$ is the electric charge in the gauge theory, which is related
to the magnetic charge by the quantisation
condition~\footnote{Equation~(\ref{ggtilde}) is defined up to an
  overall constant, which is fixed by stability of the
  monopole~\cite{Brandt:1979kk}, and irrelevant for the present
  argument.}
\begin{equation}
g\sim\frac{1}{\tilde{g}}.
\label{ggtilde}
\end{equation}
Choosing the specific transformation matrix~\footnote{Our presentation
  differs from that of ref.~\cite{Brandt:1979kk} due to our choice of
  Hermitian, rather than anti-Hermitian, generators.}
\begin{equation}
{\bf U}=e^{i\phi\sigma_3}e^{i\theta\sigma_2}e^{-i\phi\sigma_3},
\label{Udef}
\end{equation}
one finds that the transformed field in spherical polar coordinates is
given by 
\begin{equation}
{\bf A}'_\mu=\left(0,0,\frac{\tilde{g}}{r}(\sin\phi \,{\bf \sigma}_1
-\cos\phi\,{\bf \sigma}_2),
\frac{\tilde{g}}{r}(\cos\theta\cos\phi\,{\bf \sigma}_1+\cos\theta
\sin\phi\,{\bf \sigma}_2-\sin\theta\,{\bf \sigma}_3)\right)_{\rm spherical}.
\label{Aprimedef}
\end{equation}
Finally, transforming back to Cartesian coordinates, one finds 
\begin{equation}
{A'}_0^a=0,\quad {A'}^a_i=-\frac{\tilde{g}\epsilon_{iak} x_k}{r^2},
\label{AWY}
\end{equation}
such that using eq.~(\ref{ggtilde}) returns precisely the Wu-Yang form
of eq.~(\ref{WuYang}). This is distinctly different from the
abelian-like form of eq.~(\ref{ADform}): the Levi-Cevita symbol mixes
the spatial and gauge indices. In addition, the form of
eq.~(\ref{AWY}) no longer linearises the Yang-Mills equations, thus is
genuinely non-abelian in this gauge~\footnote{The fact that one can
  gauge away the non-abelian nature of the monopole may be related to
  its static nature~\cite{Sikivie:1978sa}.}. \\

This result immediately explains why the analysis of
section~\ref{sec:ultraboost} failed to find conclusive evidence that
the ultraboosted biadjoint monopole could be potentially matched with
a Wu-Yang monopole: the Wu-Yang monopole is nothing other than the
abelian-like Dirac monopole in disguise. In the Kerr-Schild double
copy, the latter is known to be related to the NUT solution in
gravity~\cite{Luna:2015paa}. The ultraboosted Dirac monopole then
double copies to a so-called NUT wave~\cite{Argurio:2008nb}. We may
confirm this explicitly, by taking the gauge field of
eq.~(\ref{Amatrix}) to be static in the inertial frame $S'$ of
section~\ref{sec:ultraboost}, and boosting to the frame $S$ of
eq.~(\ref{Lorentz}). The result is
\begin{equation}
{A}^a_\mu=\frac{\tilde{g}\delta^{a 3}}
{[\gamma^2(x-\beta t)^2+\rho^2]^{1/2}
\{[\gamma^2(x-\beta t)^2+\rho^2]^{1/2}+z\}}
\left(\begin{array}{c}\gamma\beta y\\-\gamma y\\\gamma(x-\beta t)
\\0\end{array}\right).
\label{Diracboost}
\end{equation}
Taking the limit $\gamma\rightarrow\infty$ and using
eq.~(\ref{deltalim}), one finds
\begin{equation}
A^a_\mu=\delta^{a3}\phi_M \bar{k}_\mu,
\label{Diracboost2}
\end{equation}
where $\bar{k}_\mu$ has been given in eq.~(\ref{barkmu}), and 
\begin{equation}
\phi_M=\tilde{g}\left[\pi-2\tan^{-1}\left(\frac{z}{y}\right)\right]\delta(u).
\label{phiMdef}
\end{equation}
This has the form of a delta function in the lightcone coordinate $u$,
dressed by a profile function $\phi_M$ whose dependence is only on the
coordinates in the transverse plane, thus is indeed a shockwave. Note
that the function $\phi_M$ has a cut in the $(y,z)$ plane, that is a
remnant of the original Dirac string. As for the latter, we may choose
where to place this cut in the shockwave plane, by performing gauge
transformations. This would affect the constant term in
eq.~(\ref{phiMdef}), which is therefore not physical by itself. \\

We may also note that the function $\phi_M$ is harmonic, and that
$\bar{k}_\mu$ (as has already been pointed out above) is null and
geodesic. Following the rules of ref.~\cite{Monteiro:2014cda}, we are
thus entitled to take the double copy of eq.~(\ref{Diracboost2}) to
obtain the graviton
\begin{equation}
h_{\mu\nu}=N\left[\pi-2\tan^{-1}\left(\frac{z}{y}\right)\right]\delta(u) \,
\bar{k}_\mu\,\bar{k}_\nu,
\label{NUTwave}
\end{equation}
where we have replace the magnetic monopole charge $\tilde{g}$ with a
gravitational charge $N$. Upon identifying the latter with NUT charge,
eq.~(\ref{NUTwave}) agrees with the NUT wave solution first derived in
ref.~\cite{Argurio:2008nb}~\footnote{In order to compare with
  ref.~\cite{Argurio:2008nb}, one must change the boost direction to
  the $+z$ direction, and also move the position of the cut of the
  function $\phi_M$ in the transverse plane.}. We may also take the
zeroth copy of eq.~(\ref{Diracboost2}), to obtain a biadjoint scalar
field
\begin{equation}
\Phi^{aa'}=\lambda c^a\tilde{c}^{a'}
\left[\pi-2\tan^{-1}\left(\frac{z}{y}\right)\right]\delta(u).
\label{biadjointnut}
\end{equation}
This is clearly non-zero, and thus not at all identifiable with the
ultraboost of the non-perturbative monopole of eq.~(\ref{monopole}),
which effectively vanished!\\

\begin{figure}
\begin{center}
\scalebox{0.6}{\includegraphics{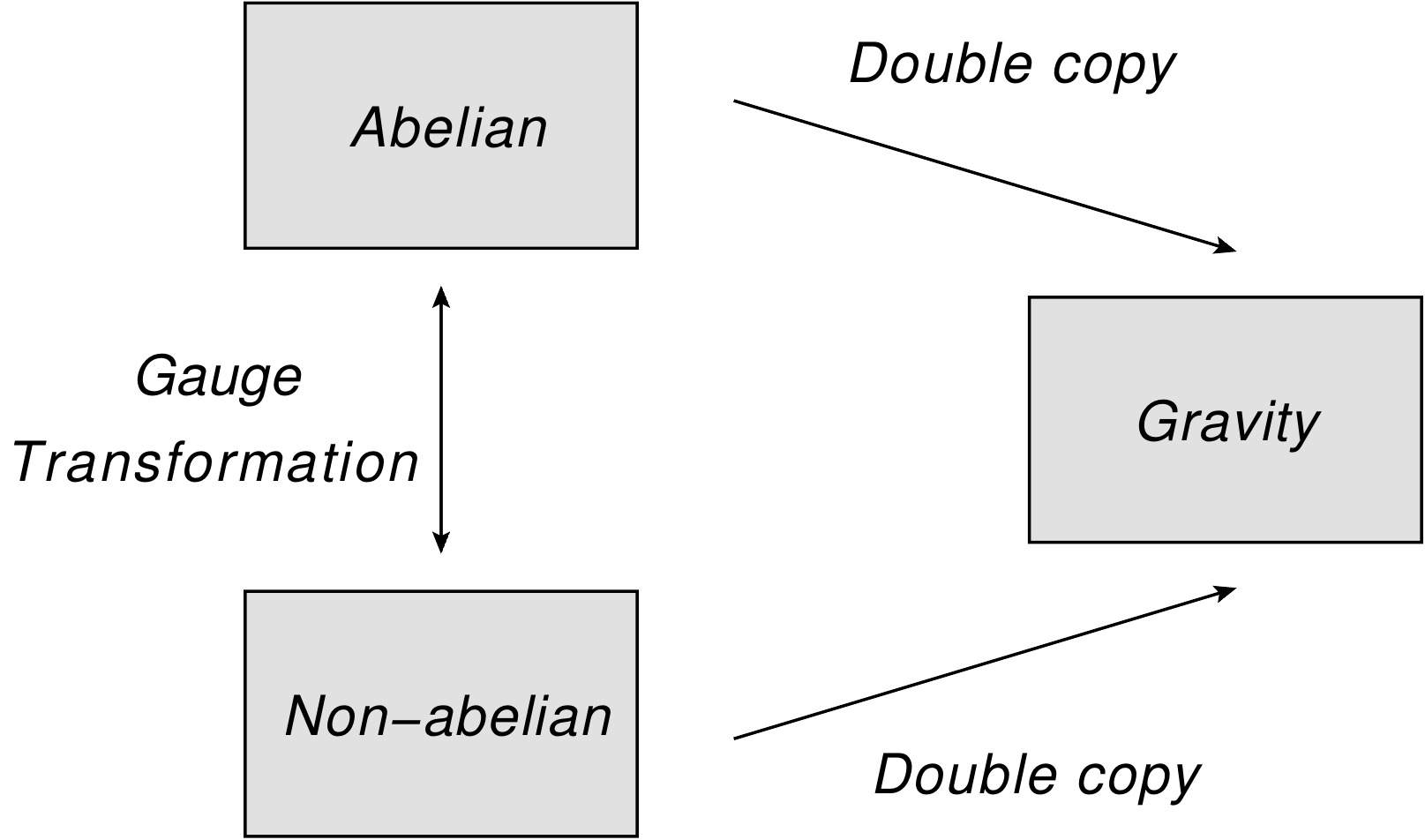}}
\caption{Generalisation of the Kerr-Schild double copy, in which one
  may identify abelian or non-abelian exact solutions of a gauge
  theory with the same gravity solution.}
\label{fig:gaugecopy}
\end{center}
\end{figure}

The results of this section are of interest in extending the remit of
the classical double copy. A frequently encountered question amongst
newcomers to the subject is why the Kerr-Schild double copy should
always produce an abelian-like solution, when the amplitude double
copy crucially relies on the fact that the gauge theory is non-abelian
(i.e. through BCJ duality~\cite{Bern:2008qj}). The natural
interpretation of the above results, however, is that one may identify
the NUT solution in gravity as a double copy either of the
abelian-like magnetic monopole of eq.~(\ref{Amatrix}), or the fully
non-abelian Wu-Yang monopole of eq.~(\ref{WuYang}). That this is
possible is due to the fact that colour information is stripped off
when taking the double copy, and the overall scheme is as depicted in
figure~\ref{fig:gaugecopy}.
Interestingly, a similar picture already has a precedent in the study
of scattering amplitudes: ref.~\cite{Oxburgh:2012zr} examined the
infrared singularities of amplitudes in either QED or QCD, and showed
that these both matched up with the same structure of IR singularities
in GR, to all orders in perturbation theory. Furthermore, it is
certainly the case that non-abelian classical solutions can be
constructed perturbatively, and double
copied~\cite{Goldberger:2016iau,Goldberger:2017frp,Goldberger:2017ogt}. However,
we believe that our results constitute the first example of an {\it
  exact} non-abelian solution that can be double copied to a
gravitational counterpart, and the extension of this to other gauge
groups and solutions deserves further investigation.\\

\section{Conclusion}
\label{sec:discuss}

In this paper, we have systematically investigated the possibility of
a non-perturbative double copy between gauge and gravity
theories. More specifically, we have followed the suggestion that the
non-perturbative biadjoint monopole of ref.~\cite{White:2016jzc} may
be related to the singular Wu-Yang monopole solution in SU(2) gauge
theory. There is no existing guidance on how to proceed, given that in
all previous examples of the double copy, solutions of the linearised
biadjoint field equation play a crucial role. For amplitudes, these
are the denominator factors that act as scalar propagators. For the
Kerr-Schild double copy, one relies upon a harmonic function $\phi$,
that remains untouched (up to replacements of charge and mass factors)
when moving between theories. \\

Our starting point was to note that shockwaves are known to
double-copy, and that one can construct such solutions by
ultraboosting static objects, by analogy with the seminal gravity
study of ref.~\cite{Aichelburg:1970dh}. Furthermore, ultraboosting a
solution results in a somewhat simpler structure, which might make it
easier to spot how a nonperturbative double copy might work. To this
end, we first showed how the Aichelburg-Sexl procedure, for the point
charge and mass solutions considered in the original classical double
copy of ref.~\cite{Monteiro:2014cda}, can be expressed in a form that
makes the double copy manifest. This exercise was useful in itself:
the ultraboost procedure has not been previously considered in
biadjoint theory, and there are interesting aspects of how the double
copy relates the ultraboosts in different theories (e.g. the need to
rescale the charge parameters differently). Returning to the biadjoint
monopole, we found that this effectively vanished upon performing an
ultraboost, making any attempt to match it up with a gauge theory
solution difficult.\\

We then explained the above observations by noting that the Wu-Yang
monopole is related to a trivially dressed Dirac monopole by a
(singular) gauge transformation. The latter, as implied by the study
of ref.~\cite{Luna:2015paa}, maps to the well-known NUT solution in
gravity. Consequently, upon ultraboosting the transformed Wu-Yang
monopole, one may double copy the result to obtain the so-called NUT
wave of ref.~\cite{Argurio:2008nb}, which we verified by explicit
calculation. Interestingly, we expressed the Dirac monopole in a gauge
that was not in the right form for the usual Kerr-Schild double copy
(e.g. the field was not null). However, the shockwaves we obtained
were indeed of Kerr-Schild form, which may possibly be due to the
highly constrained symmetry of the shockwave solutions. In addition,
this then allowed us to take the zeroth copy of the boosted Dirac
monopole, which is non-zero and thus not relatable to the biadjoint
monopoles of ref.~\cite{White:2016jzc}.\\

Our results provide a number of insights at the frontier of
understanding of the remit of the double copy. Most significantly, we
have uncovered the first example of an exact non-abelian solution -
the Wu-Yang monopole - that can be identified with the same gravity
solution as an abelian-like counterpart. This is analogous to similar
behaviour observed in the study of scattering
amplitudes~\cite{Oxburgh:2012zr}, and indeed suggests that many more
such examples can be found. This may provide clues about how to double
copy exact non-abelian solutions without having to rely on the
Kerr-Schild procedure, which would in turn greatly increase our
understanding of the fundamental origin of the double copy itself.

\section*{Acknowledgments}

We thank Donal O'Connell for a discussion, some time ago, about
ultraboosting classical solutions, and David Berman for further useful
discussions. NBA and RSM are supported by PhD studentships from the
United Kingdom Science and Technology Facilities Council (STFC) and
the Royal Society respectively. This work has been supported by the
STFC Consolidated Grant ST/P000754/1 ``String theory, gauge theory and
duality'', and by the European Union Horizon 2020 research and
innovation programme under the Marie Sk\l{}odowska-Curie grant
agreement No. 764850 ``SAGEX''.

\appendix
\section{Impulse Calculations} \label{app:impulse}
\subsection{Biadjoint Impulse}
The impulse of a test particle of mass $m$ interacting with the
ultraboosted biadjoint scalar field is related to the equations of
motion for the interacting particle coupled with the biadjoint
scalar~\cite{Goldberger:2017frp},
\begin{equation}
\frac{\partial p_\mu}{\partial t}=-\frac{\lambda}{m} c^a_2 \tilde{c}^{a'}_2\partial_\mu \Phi^{aa'}.
\end{equation}
Here, the biadjoint field $\Phi^{a\dot{a}}$ is given by the shockwave solution given by (\ref{Phiboosted}). Stripping out charge and mass parameters for the sake of simplicity, we have
\begin{equation}
\frac{\partial p_\mu}{\partial t}=\partial_\mu \left( \ln\left(\frac{\rho^2}{\rho_0^2}\right)\delta(x-t)\right).
\end{equation}
Switching to lightcone coordinates $u=t-x$, $v=t+x$, the impulse
experienced by the particle is
\begin{equation}
\delta p_\mu=\int^\infty_{-\infty} \ dt \ \frac{\partial p_\mu}{\partial t}
=\int^\infty_{-\infty} \ du \ \frac{\partial p_\mu}{\partial u}
=\int^\infty_{-\infty} \ du \ \partial_\mu \left( \ln\left(\frac{\rho^2}{\rho_0^2}\right)\delta(u)\right).
\end{equation}
We will work this out for each component,
\begin{equation*}
\delta p_u =\int^\infty_{-\infty} du \partial_u \left( \ln\left(\frac{\rho^2}{\rho_0^2}\right)\delta(u)\right)
= \ln\left(\frac{\rho^2}{\rho_0^2}\right)\delta(u)\bigg|_{u=-\infty}^{\infty}=0
\end{equation*} 
\vspace{-10pt}
\begin{align*}
\delta p_v =0, && 
\delta p_y =\int^\infty_{-\infty} \ du \ \frac{2y}{\rho^2} \ \delta(u)=\frac{2y}{\rho^2}, &&
\delta p_z =\int^\infty_{-\infty} \ du \ \frac{2z}{\rho^2} \ \delta(u)=\frac{2z}{\rho^2}.
\end{align*}
Altogether, the interacting particle experiences an impulse with charges and couplings reinstated:
\begin{equation}
\delta p^\mu=-\frac{\lambda^2}{m}\frac{c\cdot c_2\,\tilde{c}\cdot\tilde{c}_2}{2 \pi \rho^2} \left(0,0,y,z \right).
\end{equation}

\subsection{Gauge Theory Impulse}
The impulse of a particle interacting with the ultraboosted gauge field is related to the Lorentz force, given by
\begin{equation}
\frac{\partial p^\mu}{\partial t}=g c^a_2  F^{ a \mu}{}_{\nu} v^\nu.
\end{equation}
As we are working in the rest frame of the interacting particle, we only need to consider the following components of the field strength tensor:
\begin{equation}
F^{a\mu{}}_0=\partial^\mu A^a_0-\partial_0 A^{a\mu} 
\end{equation}
where  $A^\mu$ is the shockwave gauge potential given by the ultraboosted field given by (\ref{Aboostedres}), as well as $A_\mu =A^{a}_{\mu}T^a $, and $\partial^\mu=(\partial_t,-\partial_x,-\partial_y, -\partial_z)$. The components of interest of the field strength tensor are
\begin{align}
F^{a1}{}_0&=\frac{gc^a}{4 \pi}\ln\left(\frac{\rho^2}{\rho_0^2}\right)(\partial_t+\partial_x)\delta(x-t)=0
\notag \\
F^{aj}{}_0&=\frac{gc^a}{4 \pi}\delta(x-t)\partial_j \ln\left(\frac{\rho^2}{\rho_0^2}\right)
=\frac{c^a}{4 \pi}\delta(x-t) \ \frac{2x^j}{\rho^2}
\end{align}
where $j=2,3$ or equivalently $x^j=y,z$. We can now extract the impulse,
\begin{equation}
\delta p^\mu=\int^\infty_{-\infty} \ dt \ \frac{dp^\mu}{dt} \ \ \rightarrow
\delta p^j=\int^\infty_{-\infty} \ dt \ \frac{c_2 \cdot c}{4 \pi}\delta(x-t) \frac{2x^j}{\rho^2} 
=\frac{c_2 \cdot c}{4 \pi}\frac{2x^j}{\rho^2} 
\end{equation}
which leaves us finally with the impulse
\begin{equation}
\delta p^\mu:\frac{g^2 c_2 \cdot c}{2 \pi \rho^2}(0,0,y,z).
\label{gaugekick}
\end{equation}
This is identical (setting aside couplings and charges) to the biadjoint case. If the interacting particle were to have the same charge as that producing the shockwave, the effect of the impulse would be to send the particle away from the shockwave nucleus. Opposite charges would result in the particle being drawn in closer to the nucleus. The strength of the push or pull depends on how far the particle is initially from the source of the shockwave. The closer the initial separation, the stronger the impulse.
\subsection{Gravitational Impulse}
The impulse of a particle of mass $m$ interacting with the ultraboosted graviton field is related to the geodesic equation given by
\begin{equation}
\frac{\partial p^\mu}{\partial t}=-m\Gamma^\mu_{\; \nu \sigma} v^\nu  v^\sigma,
\end{equation}
The metric associated with shockwave geometry is
\begin{equation}
g_{\mu \nu}=\eta_{\mu \nu}+\frac{\kappa^2}{2}\frac{M}{4 \pi}\delta(x-t) \ln \left(\frac{\rho^2}{\rho_0^2}\right) \bar{k}_\mu \bar{k}_\nu
\end{equation}
with inverse
\begin{equation}
g^{\mu \nu}=\eta^{\mu \nu}-\frac{\kappa^2}{2}\frac{M}{4 \pi}\delta(x-t) \ln \left(\frac{\rho^2}{\rho_0^2}\right) \bar{k}^\mu \bar{k}^\nu.
\end{equation}
As we are in the rest frame of the particle interacting with our shockwave, the geodesic equation simplifies greatly to
\begin{equation}
\frac{\partial p^\mu}{\partial t}=-m\Gamma^\mu_{\;00}.  
\end{equation}
We only need 4 Christoffel symbols:
\begin{align*}
\Gamma^0_{\;00} =\frac{-1}{2}\frac{\kappa^2}{2}\frac{M}{4 \pi}\ln\left(\frac{\rho^2}{\rho_0^2}\right)\partial_t\delta(x-t), &&
\Gamma^x_{\;00} =\frac{1}{2}\frac{\kappa^2}{2}\frac{M}{4 \pi}\ln\left(\frac{\rho^2}{\rho_0^2}\right)\partial_t\delta(x-t) ,
\end{align*}
\vspace{-10pt}
\begin{align*}
\Gamma^y_{\;00} =\frac{\kappa^2}{2}\frac{M}{4 \pi} \frac{y}{\rho^2} \delta(x-t), &&
\Gamma^z_{\;00} =\frac{\kappa^2}{2}\frac{M}{4 \pi} \frac{z}{\rho^2} \delta(x-t).
\end{align*}
The impulse $\delta p^\mu$ is then the time integral of these Christoffel symbols,
\begin{align}
\delta p^0= -\delta p^x &=\frac{\kappa^2}{2}\frac{Mm}{4 \pi}\int^\infty_{-\infty}  dt \frac{1}{2}\ln\left(\frac{\rho^2}{\rho_0^2}\right) \partial_t \delta(x-t) \notag \\[5pt]
&=\frac{\kappa^2}{4}\frac{Mm}{4 \pi}\ln\left(\frac{\rho^2}{\rho_0^2}\right)\int^\infty_{-\infty}  du \ \partial_u\delta(u) \notag \\[5pt]
&=\frac{\kappa^2}{4}\frac{Mm}{4 \pi}\ln\left(\frac{\rho^2}{\rho_0^2}\right) \delta(u)\bigg|_{u=-\infty}^{\infty}=0 \notag \\[10pt]
\delta p^y &=\frac{-\kappa^2}{2}\frac{Mm}{4 \pi}\int^\infty_{-\infty}  dt \ \delta(x-t) \frac{y}{\rho^2} =\frac{-\kappa^2}{2}\frac{Mm}{4 \pi}\frac{y}{\rho^2} \notag \\[5pt]
\delta p^z &=\frac{-\kappa^2}{2}\frac{Mm}{4 \pi}\int^\infty_{-\infty}  dt \ \delta(x-t) \frac{z}{\rho^2} =\frac{-\kappa^2}{2}\frac{Mm}{4 \pi}\frac{z}{\rho^2}.
\end{align}
The impulse bears a resemblance to those of the biadjoint scalar and gauge theories, i.e.
\begin{equation}
\delta p^\mu=\frac{-\kappa^2 Mm}{8 \pi \rho^2}(0,0,y,z).
\label{gravkick}
\end{equation}
The particle gets a kick restricted to the $y-z$ plane, with the
magnitude of the kick diminishing the further the particle is from the
nucleus of the shockwave. We would expect, as gravity is an attractive
force, that the impulse should draw the particle toward the nucleus of
the shockwave. By contrast, the gauge theory case is repulsive
(provided $c_2\cdot c$ is positive in eq.~(\ref{gaugekick})), which
results in the sign difference between eq.~(\ref{gaugekick}) and
eq.~(\ref{gravkick}).

\section{Ultraboosting the general SU(2)$\times$SU(2) Monopole}
\label{app:generalmonopole}

In section~\ref{sec:biadointboost}, we ultraboosted the biadjoint
monopole solution of eq.~(\ref{monopole}), finding that it disappears
in the ultrarelativistic limit. In this appendix, we show that a
similar conclusion is reached for the more general monopole solutions
of eq.~(\ref{monopole2}), that were are also derived in
ref.~\cite{White:2016jzc}. Recall that these solutions have the form
\begin{equation*}
\Phi^{aa'}=\frac{1}{\lambda r^2}\left[-k\left(\delta^{aa'}-\frac{x^a x^{a'}}
{r^2}\right)\pm\sqrt{2k-k^2}\frac{\epsilon^{aa'd}x^d}{r}\right],\quad
0\leq k\leq 2.
\end{equation*}
Intermediate details of the
ultraboost calculation are cumbersome due to selecting a
particular boost direction, so we will simply quote the results. As in
section~\ref{sec:biadointboost}, we boost in the $x$-direction, and
examine the behaviour of the boosted solution inside and outside the
plane $x=\beta t$. Outside of the plane, we find that the solution
vanishes completely:
\begin{align*}
\Phi^{aa'} \xrightarrow{\gamma \rightarrow \infty} 0 && \text{for} \;\; x-\beta t \neq 0.
\end{align*}
Inside of the plane, just as for the ultraboost of the
\eqref{monopoleboost2}, the field takes finite values and displays no
divergent behaviour. More specifically, in the limit $\gamma
\rightarrow \infty$ we find
\begin{align*}
\Phi^{11} \rightarrow \frac{-k}{\lambda (y^2 + z^2)}, && \Phi^{22} \rightarrow \frac{-kz^2}{\lambda (y^2 + z^2)^2}, && \Phi^{33} \rightarrow \frac{-ky^2}{\lambda (y^2 + z^2)^2},
\end{align*}
\begin{align*}
\Phi^{12} = -\Phi^{21} \rightarrow \frac{z\sqrt{(k^2-2k)}}{\lambda(y^2 + z^2)^{3/2}}, && \Phi^{13} = -\Phi^{31} \rightarrow -\frac{y\sqrt{(k^2-2k)}}{\lambda(y^2 + z^2)^{3/2}}, && \Phi^{23} = \Phi^{32} \rightarrow \frac{yzk}{\lambda (y^2 + z^2)^2}.
\end{align*}
We thus conclude that, similarly to \eqref{monopoleboost2}, the
solutions of eq.~(\ref{monopole2}) remain finite in the plane $x=\beta
t$ after the ultraboost (i.e. do not produce a delta function). Thus,
they do not constitute shockwaves. Similarly to \eqref{monopole}, one
can produce a shockwave solution by rescaling the coupling according
to $\lambda \rightarrow \lambda / \gamma$. As mentioned below
\eqref{phiboostrescale} however, this entails changing the theory, and
is not analogous to the Aichelburg-Sexl-like rescalings, which affect
the charges of the objects being boosted.

\bibliography{refs.bib}
\end{document}